\documentclass[graybox]{svmult}

\usepackage{type1cm}        
\usepackage{makeidx}         
\usepackage{graphicx}        
\usepackage{multicol}        
\usepackage[bottom]{footmisc}

\usepackage{newtxtext}       %
\usepackage{newtxmath}       

\newcommand{\todo}[1]{#1}
\newcommand{\stolen}[1]{#1}

\makeindex             

\begin{document}

\title*{On the Reliability of Computing-in-Memory Accelerators for Deep Neural Networks}
\author{Zheyu Yan, Xiaobo Sharon Hu, and Yiyu Shi}
\institute{Zheyu Yan \at University of Notre Dame, Notre Dame, IN, United States, \email{zyan2@nd.edu}
\and Xiaobo Sharon Hu \at University of Notre Dame, Notre Dame, IN, United States \email{shu@nd.edu}
\and Yiyu Shi \at University of Notre Dame, Notre Dame, IN, United States \email{yshi4@nd.edu}}
%
%
\maketitle


\abstract{Computing-in-memory with emerging non-volatile memory (nvCiM) is shown to be a promising candidate for accelerating deep neural networks (DNNs) with high energy efficiency. However, most non-volatile memory (NVM) devices suffer from reliability issues, resulting in a difference between actual data involved in the nvCiM computation and the weight value trained in the data center. Thus, models actually deployed on nvCiM platforms achieve lower accuracy than their counterparts trained on the conventional hardware (\emph{e.g.}, GPUs). In this chapter, we first offer a brief introduction to the opportunities and challenges of nvCiM DNN accelerators and then show the properties of different types of NVM devices. We then introduce the general architecture of nvCiM DNN accelerators. After that, we discuss the source of unreliability and how to efficiently model their impact. Finally, we introduce representative works that mitigate the impact of device variations.}

\section{Introduction}
Deep Neural Networks (DNNs) have excelled human performance in various crucial tasks (\emph{e.g.}, image classification, object detection, and speech recognition) and have become a popular solution for them. Thus, edge devices such as automobiles, smartphones, and smart sensors that depend on these tasks are ideal platforms to be empowered by DNNs. However, due to the constrained computation resource and limited power budget of edge devices, direct implementation of computational intensive DNNs on edge devices is a significant challenge. 


A majority of the works addressing this challenge use application-specific integrated circuits (ASICs) or field-programmable gate arrays (FPGAs) for DNN acceleration. These conventional special-purpose edge DNN accelerators typically use a group of on-chip process elements (PEs) to handle computation and utilize off-chip non-volatile (NV) storage (\emph{e.g.}, flash) to store the model information (\emph{i.e.}, DNN architecture and model weights)~\cite{chen2019cmos}. Between the static random-access memory (SRAM) in the PEs used for temporary data caching and off-chip non-volatile storage used for power-off data preservation, there is also a complex memory hierarchy, generally consisting of several levels of dynamic random access memory (DRAM)-based on-chip memories. Because of this separation of data and computation, which is a key limitation of the conventional von-Neumann architecture, these DNN edge accelerators face energy efficiency and computing latency challenges. Specifically, PEs of this kind of architecture generates a large volume of intermediate data. These intermediate data need to be moved between different levels of the memory hierarchy so that they can be used by different process elements. Data movements across different levels of memory hierarchy induce a great time and energy consumption overhead, especially when accessing the lower level of the memory hierarchy. This challenge is also called \emph{the memory wall}.

Non-volatile Computing-in-Memory (nvCiM) DNN accelerators~\cite{ielmini2018memory} offer a great opportunity to break the memory wall by utilizing their special architectural advantages. nvCiM architectures reduce data movement with an in-situ weight data access scheme~\cite{sze2017efficient}. \todo{Emerging NVM devices (\emph{e.g.}, RRAMs, STT-RAMS, and FeFETs) are utilized so that nvCiM platforms can achieve higher energy efficiency and memory density compared with traditional MOSFET~\cite{shafiee2016isaac} based designs.} More specifically, nvCiM can achieve low latency and high energy efficiency because, (1) the CiM structure avoids the long latency for moving data across multi-level memory hierarchies to retrieve the intermediate data and/or DNN weights; (2) analog computing engine performs dot-product in a compact manner, thus reducing the amount of intermediate data (\emph{i.e.}, partial sums) generated in multiply-and-accumulate (MAC) operations; (3) the crossbar structured matrix-vector multiplication (VMM) engine offers high parallelism that can perform VMM in one CiM cycle, thus shortening the latency of DNN operations.

However, such accelerators suffer greatly from design limitations. \todo{Firstly, because of emerging NVM devices tend to have low precision (1-4 bits) of  (\emph{i.e.}, single NVM device can only represent 1 to 4-bit data and more than one device is needed to represent data in higher precision) and the limit of chip area, weight precision of neural networks mapped to nvCiM accelerators is limited. Secondly, most nvCiM accelerators require digital-to-analog converters (DACs) to convert the digital input data to analog signals so that it can be processed in crossbars and also analog-to-digital converters (ADCs) to convert the computation results back to digital signal for other neural network operations (\emph{e.g.}, activation and normalization). The precision of intermediate activation data is limited by the precision of DACs and ADCs. Thirdly, NVM devices, ADCs, and DACs suffer from device-to-device variations due to manufacture and programming defects and cycle-to-cycle variations due to the computational environment difference. Compared with their digital counterparts that can tolerate such noises, because of the analog nature of nvCiM accelerators, the calculations performed in these platforms are not noise-free.} The noisy nature of the computations leads to performance degradation that (1) models deployed on nvCiMs typically gets lower accuracy than their ideal counterparts trained in the data centers and (2) developers will not be able to know the exact accuracy of a model before it is deployed on a certain copy of the nvCiM product.

This reliability issue and its impact on DNN performance have been studied from different levels of design, including behavioral level explorations~\cite{yan2020single, gao2021bayesian}, architecture level analysis~\cite{yan2021uncertainty}, and device-level observations~\cite{zhao2017investigation}. Cross-layer co-design efforts that simultaneously explore DNN model and hardware design pairs that can together achieve both high perception task performance and desirable hardware reliability are the current direction of this field~\cite{jiang2020device}. 

In this chapter, we focus on design efforts targeting crossbar-based nvCiM DNN accelerators. We first introduce three typical emerging NVM devices including resistive random access memory (RRAM), ferroelectric field-effect transistor (FeFET), and Spintronics (STT) Devices. We then describe typical nvCiM DNN accelerator designs, their key components, and their benefits. After that, we discuss the limitations of nvCiM DNN accelerators and some key findings for these limitations. Finally, we introduce methods proposed to address the unreliability issue of nvCiM DNN accelerators from three aspects, encoding, DNN model training, and DNN architecture selection.
\section{Non-Volatile Devices}

\subsection{RRAM}


Resistive random access memory (RRAM) is a two-terminal device that can be programmed into different levels of resistance value by using programming voltages in different magnitudes and duration. 

\begin{figure}[h]
\begin{center}
\centerline{\includegraphics[trim=0 0 00 0, clip, width=1\linewidth] {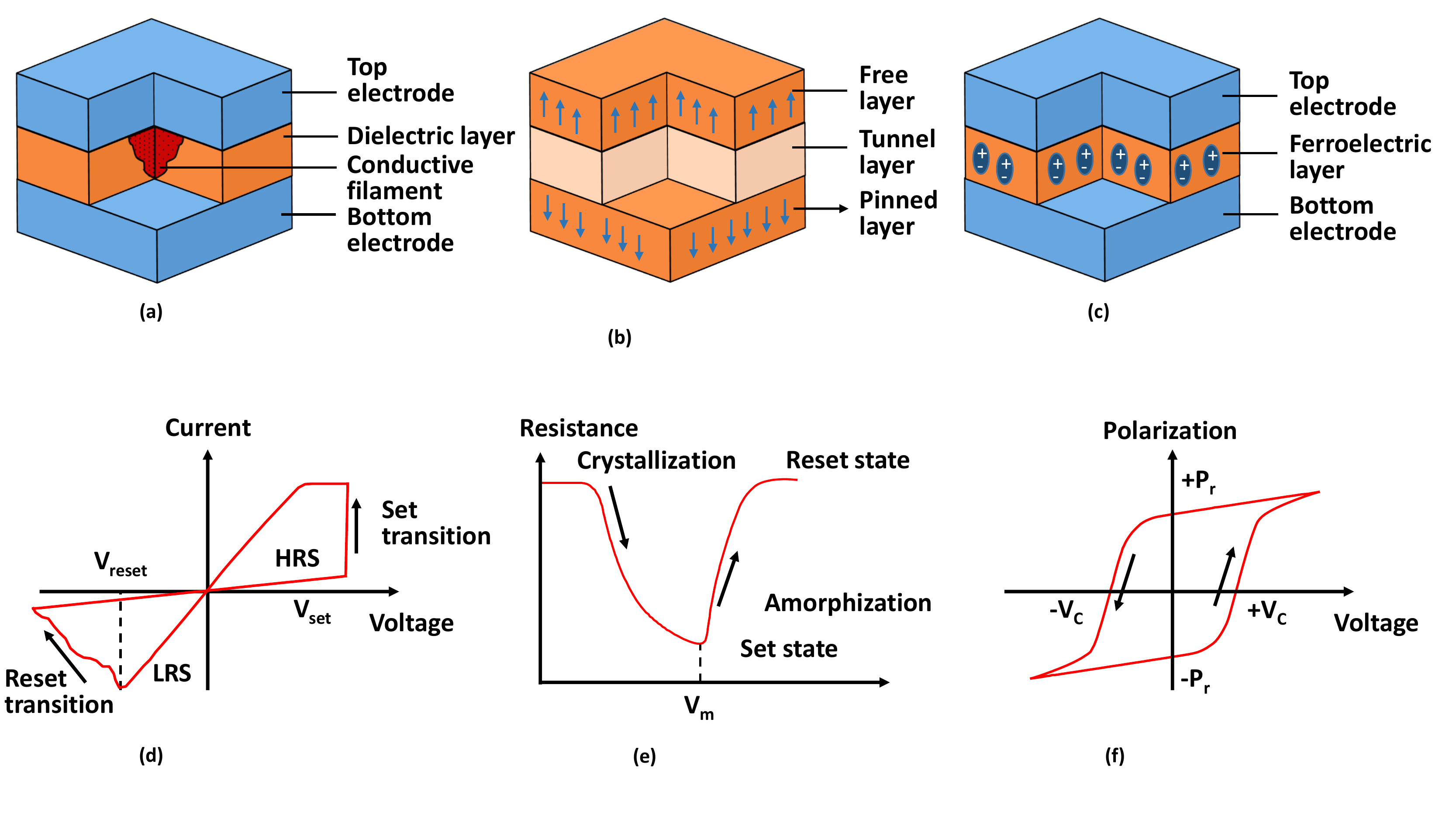}}
\caption{Illustrations for key structures of different emerging NVM devices and their characteristics~\cite{chen2019cmos}. (a) RRAMs, (b) MRAMs and (c) FeFETs. The actual devices are more complex then these illustrations. (d-f) are their characteristics, respectively.}
\label{fig:fets}
\end{center}
\end{figure}

\stolen{As shown in Fig.~\ref{fig:fets}~(a), the major component of RRAMs is a metal-insulator-metal (MIM) stack, where a dielectric layer is stacked in the middle of two electrode layers. When provided a programming voltage, a filamentary path, also called conductive filament (CF)~\cite{ielmini2011modeling}, is created by soft electrical breakdown or forming in the electrode layers. In this filamentary path, a large concentration of defects, \emph{e.g.}, oxygen vacancies in metal oxides~\cite{beck2000reproducible} or metallic ions injected from the electrodes~\cite{liu2012real}, are then driven by field-induced migration and diffusion. Application of a positive voltage to the top electrode, where the defects are concentrated, induces defect migration towards the bottom electrode, thus causing the transition to the low-resistance state (LRS), because conduction is enhanced at defect sites. Application of a negative voltage, to the contrary, induces defect migration back to the top electrode, thus causing the transition to the high-resistance state (HRS) due to the disconnection of the CF. These transitions can be seen in the idealized current-voltage (I-V) characteristic in Fig.~\ref{fig:fets}~(d), where the transition to the LRS (set operation) and the transition to the HRS (reset operation) occur at opposite voltages. Similar to the bipolar RRAM concept shown in Fig.~\ref{fig:fets}~(d), unipolar RRAMs have also been presented, where the set and reset processes both occur under the same voltage polarity because of the dominant role of Joule heating in creating and dissolving the CF~\cite{yang2013memristive, kim2011nanofilamentary}. All of these devices rely on the diffusion and migration of defects and will be referred to as RRAM throughout this chapter.}

\stolen{RRAM is a promising technology for in-memory computing thanks to the key features discussed below. First, its resistance ratio between HRS and LRS (on/off ratio) is generally greater than ten, which allows a clear distinction between digital ‘0’ and ‘1’. This feature can be further exploited by dividing this gap between HRS and LRS in a non-binary manner, \emph{i.e.}, into multiple levels, resulting in a multi-level device that can represent multiple bits of data. This helps RRAM to offer a high-density storage scheme. Secondly, RRAM can operate at a moderately high switching speed (typically below 100 ns and some devices can achieve even in the sub-ns regime~\cite{choi2016high, loke2012breaking}). Thus, RRAMs can operate in platforms with high clock speeds. Finally, RRAM is more durable compared to conventional flash storage devices~\cite{lee2011fast}. This makes training DNNs on RRAM-based platforms possible.}

\subsection{Spintronics Devices}

The spintronics devices are two or three-terminal devices equipped with a magnetic tunnel junction (MTJ) that stores information using magnetization direction of its recording layer and utilizes tunnel magnetoresistance (TMR) effect for reading where the resistance of the MTJ changes according to the stored information. In this section, we introduce the two-terminal version of this device, named spin-transfer torque (STT) device. When used as a programmable memory, this kind of device is also called Spin-transfer torque magnetic RAM (STT-MRAM).

\stolen{Fig.~\ref{fig:fets}~(b) shows a magnetic tunnel junction (MTJ), which is the major building block for most Spintronics Devices. The MTJ consists of a MIM structure where two ferromagnetic metal layers are divided by a thin tunnel oxide. An example of ferromagnetic metal materials used in MTJs is the CoFeB alloy, and an example material for the tunnel oxide is MgO. For the two ferromagnetic layers, one is referred to as the pinned layer and the other as the free layer. The magnetic polarization of the pinned layer is structurally fixed so that it can act as a reference point. On the other hand, the magnetic polarization of the free layer can be modified by a programming procedure.}

\stolen{Depending on the state of the free layer, the two ferromagnetic polarization can thus be either to the same direction (parallel) or to the opposite direction (antiparallel). Parallel polarization of the two layers puts the device into a low resistance state (LRS), and antiparallel means a high resistance state (HRS) due to the tunnel magnetoresistive effect~\cite{chappert2010emergence}. Researchers are working on finding efficient ways to flip the state of the MTJ and the spin-transfer torque (STT) is one of the newer and more competitive candidates to offer a scalable and low-efficient flip~\cite{locatelli2014spin}. In the STT procedure, transition to the parallel state takes place directly by conduction electrons, which are first spin-polarized by the pinned layer, then rotate the magnetic polarization of the free layer by magnetic momentum conservation~\cite{slonczewski1996current}. Similarly, the free layer magnetization can be rotated to the antiparallel state by applying an opposite voltage (hence opposite current direction). The relative difference in resistance of the LRS and HRS, also called the magnetoresistance ratio when referring to spintronics devices, is typically around 200\%~\cite{yuasa2004giant}. STT-based devices are also fast, with a switching speed typically lower than 1 ns, and durable, with an endurance above $10^{14}$~\cite{carboni2016understanding}}.

\stolen{In STT devices, STT induced magnetization switching~\cite{slonczewski1996current, berger1996emission} is used to store data in to the device (write process). Its primitive cell has one cell transistor and one MTJ (1T1MTJ), which can achieve a relatively small cell size of ideally $6F^2$, where $F$ is the feature size of the MTJ layer. The write current passes through the tunnel barrier, as is also the case with the read current. Accordingly, the read current should be small enough so that the write event, \emph{i.e.}, magnetization switching, does not take place, and the write current should be small enough that it does not give rise to a barrier breakdown.} 

\subsection{FeFET}



A ferroelectric transistor (FeFET) is a three-terminal device equipped with a layer of ferroelectric (FE) material. It can either be configured to a steep switching mode to serve as an efficient FET or a non-volatile (NV) mode to serve as a programmable switch.

\stolen{The structure of a FeFET is similar to a regular bulk MOSFET or FinFET, except that in its gate stack, there is an additional layer of ferroelectric (FE) material. Besides this FE material, a metal layer between the FE and dielectric may or may not be included~\cite{aziz2018computing}. Designs of FE transistor structures with~\cite{li2015sub} and without~\cite{sharma2017impact} this layer both demonstrate state-of-the-art efficiencies. It is worth noting that although some FE materials (\emph{e.g.}, hafnium zirconium oxide (HZO)) are both efficient and highly compatible with CMOS processes and can thus be realized on the industrial scale, other FE materials (\emph{e.g.}, lead zirconium titanate (PZT)~\cite{aziz2016physics}) may be incompatible with CMOS processes.}

\stolen{As discussed above, FeFETs can operate in two different modes: an NV mode or a steep switching mode. Basic structures of FeFETs in these two modes are the same, except that in different configurations (\emph{e.g.}, material thickness, gate length, and width), the relative capacitance of the FE material and the underlying FET changes, resulting in different modes of operation.} In this chapter, we discuss the properties of FeFETs in the NV mode because FeFETs used in nvCiM DNN accelerators are majorly in this mode.

\stolen{NV mode of FeFETs are discovered later than its steep switching counterpart at the emergence of HZO-based FeFETs~\cite{li2017enabling}. The non-volatile property results from the hysteretic polarization ($P$) versus voltage of the FE material ($V_{FE}$) shown in Fig.~\ref{fig:fets}~(f). When the FE material is placed in series with the gate of a transistor, the hysteretic window of $P$ versus $V_{GS}$ is reduced because the MOS structure of the FET and the associated depolarization fields imposes a capacitance and the total capacitance between gate and source changes~\cite{wang2016ferroelectric}. Nevertheless, a sufficiently thick FE broadens the hysteretic window so that the hysteretic behavior is preserved and can be observed in the $I_D-V_{GS}$ transfer characteristics of this device Fig.~\ref{fig:fets}~(f). This corresponds to the non-volatile, hysteretic mode of FeFETs. In this mode of operation, at $V_{GS} = 0 V$ (\emph{i.e.}, when the supply voltage is turned off), the FeFET exhibits two stable states which correspond to positive or negative polarization retention in the FE layer. For an n-type FeFET, the device exhibits high resistance states (HRS) when $P < 0$ and low resistance states (LRS) when $P > 0$. For a p-type FeFET, it is in HRS when $P > 0$ and LRS when $P < 0$. Thus, when the FE layer is sufficiently thick, non-volatility can be embedded inside a transistor, \emph{i.e.}, FeFET can operate as an NV memory and a transistor switch at the same time.}

\section{CiM DNN Accelerators}

\subsection{Computing-in-Memory}


Conventional von-Neumann architecture is not efficient because the cost of data movements between memory and processing units is high. This issue is called \emph{the memory wall}. More seriously, the technologies for logic units are growing faster than memory cells, causing a significant gap between computation and memory access. Thus, various efforts have been made to break \emph{the memory wall} by moving the computations closer to memory. The integration of memory and computation is an evolving concept and is developing along with technological advances~\cite{gao2019eva}. We first introduce an earlier concept which is now considered near memory computing (NMC). Researchers embed processing cores into dynamic random-access memory (DRAM) modules~\cite{oskin1998active, mai2000smart, draper2002architecture} so that data can be processed in the DRAM module. This avoids sending data 
from DRAM to CPUs across the complex memory hierarchy. However, integrating DRAMs and processing units on the same chip is not beneficial if the communication cost between memory and processing units is not reduced. The concept of 3D stacking is adopted to address this issue. By stacking multiple silicons on top of each other and utilizing through-silicon-vias (TSVs) to handle inter-silicon layer communications, 3D stacking allows the processing unit to be integrated as additional layers of the stacked chip and can provide higher bandwidth compared with putting memory and logic in different chips~\cite{farmahini2015nda, zhang2014top, hsieh2016accelerating}. However, these methods do not actually use memory modules for data processing and are still sending data from memory to logic.

A step further from NMC is computing-in-memory (CiM), where processing is directly performed inside the memory array. The latency and energy efficiency requirements of edge devices greatly inspired researches in this field. The integration of processing and memory units can be done in different levels of granularity. The extremest design of CiM is that each of the memory cells is able to perform logic operations~\cite{kvatinsky2014magic}. This is referred to as fine-grained CiM. There is also a spectrum of designs between fine-grained CiM and NMC. A typical design is to empower memory arrays (of SRAM or DRAM) with processing abilities so that data can be processed inside operations inside and between memory arrays. This can be achieved by modifying the peripheral circuitry of these memory arrays. This approach is referred to as coarse-grained CiM.

The CiM concept is further evolved with the help of new advances in emerging NVM device technologies. Specifically, NVM devices including RRAMs, STT-MRAMs, and FeFET-based RAMs can offer high density, good scalability, and high power efficiency. Thus, these devices are natural replacements for SRAMs or DRAMs in CiM architectures. Various recent efforts utilize CiM-capable NVM devices instead of SRAMs or DRAMs as building blocks of either cache or main memory. One direction of research is to use NVM simultaneously as storage and logic devices by re-designing sense amplifiers so that NVM arrays can perform a subset of logic and arithmetic operations~\cite{jain2017computing, reis2018computing, li2016pinatubo}. Another direction is to use NVMs to build content-addressable memories (CAMs). CAMs can perform searches in a parallel manner, thus reducing the search time significantly. Moreover, search in CAMs requires little data movement, which leads to low energy consumption. The third direction is to use NVM devices to build DNN accelerators. These accelerators can directly execute matrix-vector multiplication inside the memory array. This saves the cost of data movements. The advances of NVM-based CiM DNN accelerators are discussed in detail in the following sections.

\subsection{Crossbar-based Vector-Matrix Multiplication Engine}

\begin{figure}[h]
\begin{center}
\centerline{\includegraphics[trim=0 150 550 0, clip, width=0.6\linewidth] {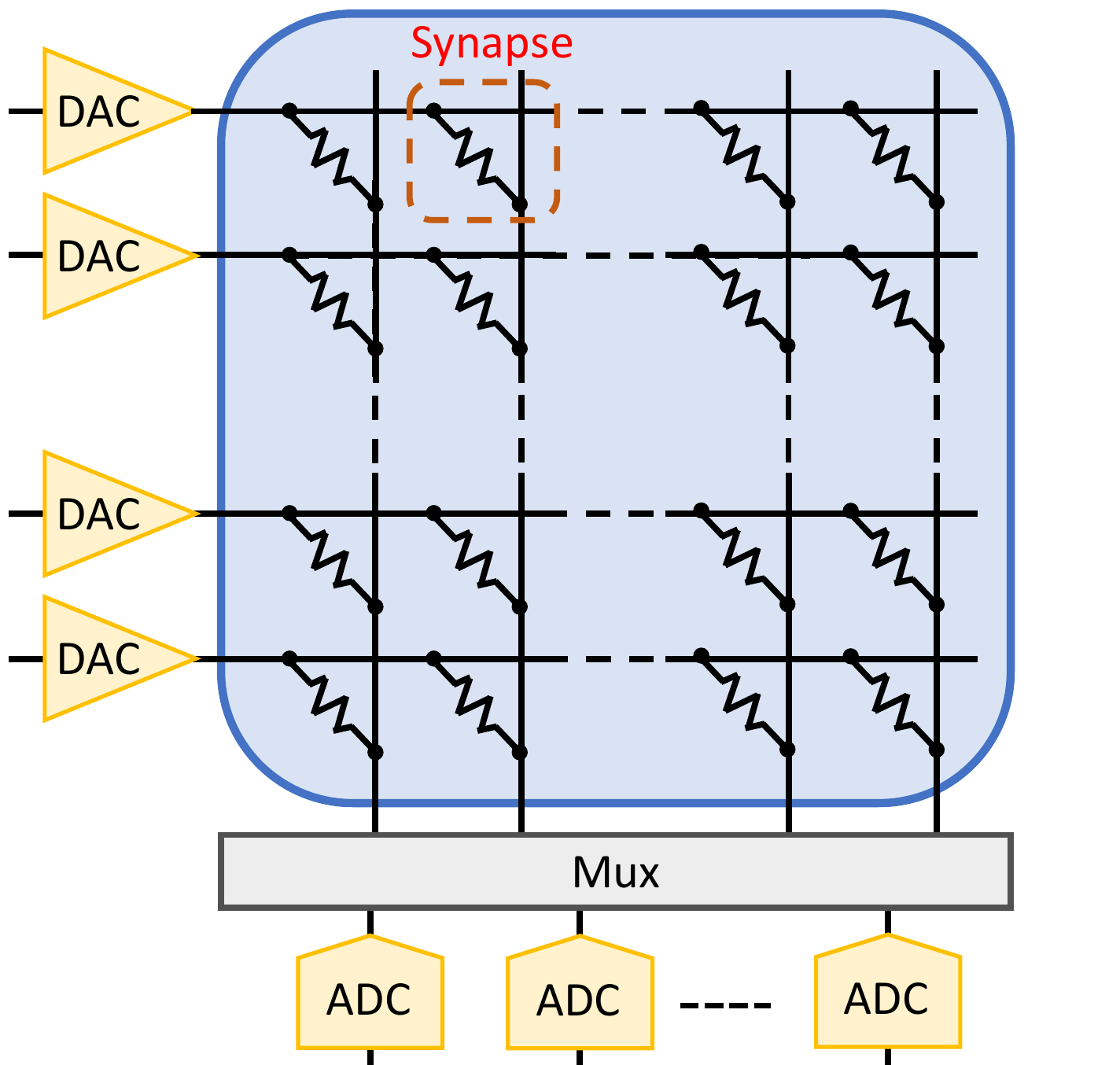}}
\caption{Illustration of crossbar array architecture. The input is fed horizontally and multiplied by weights stored in the NVM devices at each cross point. The multiplication results are summed up vertically and the sum serves as an output.}
\vspace{-0.5cm}
\label{fig:crossbar}
\end{center}
\end{figure}

Crossbar array is the key component of nvCiM DNN accelerators. As shown in Fig.~\ref{fig:crossbar}, a crossbar array can be considered as a processing element for matrix-vector multiplication where matrix value (\emph{i.e.}, weights for DNNs) are stored at the cross point of each vertical and horizontal line with resistive NVM devices such as RRAMs and FeFETs, and each vector value is propagated through horizontal data lines. In this work, we mainly introduce an RRAM-based design. Designs using other kinds of NVM devices are with similar structures. The calculation in crossbar array is performed in the analog domain but additional peripheral digital circuits are needed for other key DNN operations (\emph{e.g.}, non-linear activation and pooling), so DAC and ADCs are adopted between different components.


As is demonstrated in Fig.~\ref{fig:crossbar}, every bitline (vertical) is connected to every wordline (horizontal) via NVM cells~\cite{shafiee2016isaac}. Assume that the cells in the first column are programmed to resistances $r_1, r_2,..., r_n$, where $n$ is the number of rows. The conductances of these cells, $g_1, g_2, ..., g_n$, are the inverses of their resistances ($g_i = 1 / r_i$). If voltages $V_1, V_2,...,V_n$ are applied to each row2, cell $i$ generates current $V_i / R_i$, which is equivalent to $V_i \times g_i$, into the bitline, based on Kirchoff’s Law. The total current accumulated on the bitline is the sum of currents passed by each cell in the column, \emph{i.e.}, $I = \sum_{i=1}^{n} V_i \times g_i$. This current $I$ represents the value of a dot product operation, where one vector is the set of input voltages at each row $\mathbf{V}$ and the second vector is the set of cell conductances $\mathbf{g}$ in a column, \emph{i.e.}, $I = \mathbf{V} \cdot \mathbf{g}$.

As shown in Fig.~\ref{fig:crossbar}, $\mathbf{V}$ is applied to all columns in parallel. The currents emerging from each bitline can therefore represent multiple vector-vector dot product, which is then a vector-matrix multiplication. VMM is the key operation of DNNs. In a fully connected layer, for example, there are multiple neurons and each neuron is fed with the same input vector, but each of the neurons has a different set of synaptic weights. This operation can be represented by $\mathbf{O} = \mathbf{V} \ \mathbf{G}$ where $\mathbf{V}$ is the input, $\mathbf{G}$ is the weight matrix for neurons and $\mathbf{O}$ is its output.  The crossbar array shown in Fig.~\ref{fig:crossbar} represents an $n \times m$ crossbar array that performs dot products on $n$-entry input vectors for $m$ different outputs in a single CiM cycle.

Note that the result of the VMM operation would also need to be applied a bias value and passed through a non-linear activation function. This is done off the crossbar array. Thus, peripheral circuits are needed to perform these operations. Moreover, crossbar arrays handle VMM operations in the analog domain while other peripheral circuits are digital. DACs and ADCs are needed to transform data to and from the analogy domain. Generally, for each row of the crossbar array, there is a dedicated DAC to serve this wordline. However, ADCs are large in area and power-hungry. Thus, multiple bitlines need to share one ADC and this is achieved by the sense-and-hold circuits along with the MUX selector.

\subsection{General Architecture of nvCiM DNN accelerators}\label{sect:general_arch}


\begin{figure}[h]
\begin{center}
\centerline{\includegraphics[trim=0 80 200 0, clip, width=1\linewidth] {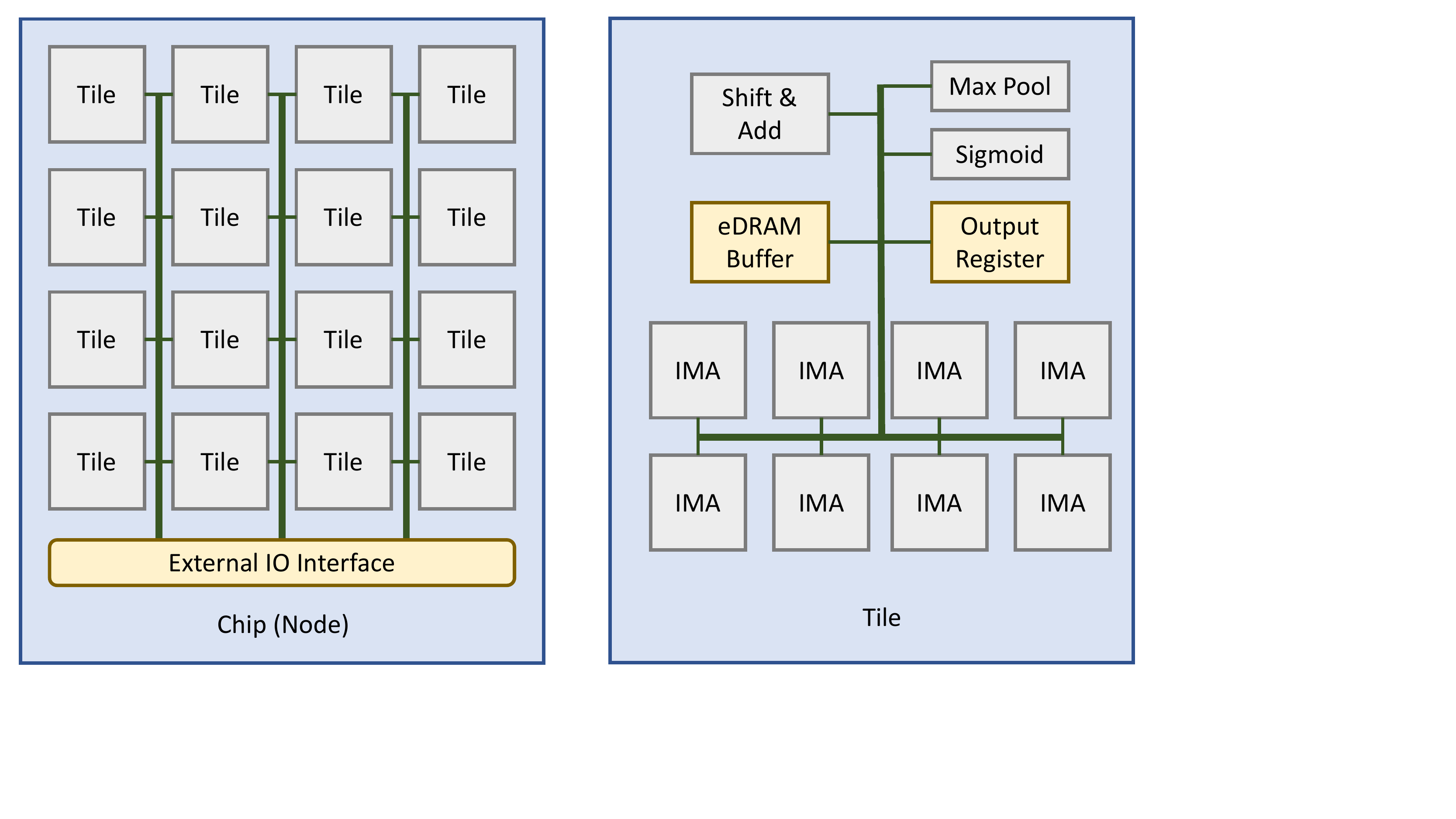}}
\caption{Illustration of ISAAC architecture. ISSAC is composed of a group of tiles and each tile consists of multiple crossbar-based IMAs, buffers, and peripheral circuits for other key DNN operations.}
\label{fig:ISAAC}
\end{center}
\end{figure}

Various accelerator architectures have been proposed to utilize the nvCiM crossbar arrays for more efficient DNN acceleration. There are generally two fashions of acceleration, one only accelerates the inference path of DNN models and the other also considers DNN training acceleration. In this chapter, we focus on DNN inference acceleration and we introduce two well-known architecture level designs, ISAAC~\cite{shafiee2016isaac} and PRIME~\cite{chi2016prime} for this scheme.

The first design, \underline{I}n-\underline{S}itu \underline{A}nalog \underline{A}rithmetic in \underline{C}rossbars (ISAAC)~\cite{shafiee2016isaac} uses crossbar arrays for both DNN weight storage and processing elements for VMM operations~\cite{zaman2021custom}. As shown in Fig.~\ref{fig:ISAAC}, ISAAC is implemented with a hierarchical-structured architecture whose major component is ``tile''. Each tile consists of multiple in-situ MAC units (IMA), eDRAM buffers, and key DNN circuitries including shift-and-add (SA), sigmoid, and max-pooling units. Thus, a tile can perform DNN operations individually. Each IMA unit is equipped with a few crossbar arrays and ADCs connected by a shared bus. Different from traditional SRAM-based designs, writing NVM devices is expensive (both in terms of time and energy consumption), so re-configuring crossbars in runtime are not feasible and thus crossbar arrays cannot be reused and each array is dedicated to only one CNN layer. The outputs of a former layer are temporarily preserved in the eDRAM buffer so that they can be used as the input of the next layer. Note that, except for the structure inside a ``tile'', the architecture of ISAAC is very similar to its digital DNN accelerator counterpart DaDianNao~\cite{chen2014dadiannao}, which is a state-of-the-art architecture when ISAAC is proposed. After tape out, the researchers show that, with a 16-chip configuration, ISAAC achieves 14.8× higher throughput while consuming 5.5× lower energy than DaDianNao. This means (1) ISAAC can achieve higher energy efficiency than state-of-the-art and (2) crossbar array-based design is a key contributor to this efficient design.

\begin{figure}[h]
\begin{center}
\centerline{\includegraphics[trim=0 50 430 0, clip, width=0.8\linewidth] {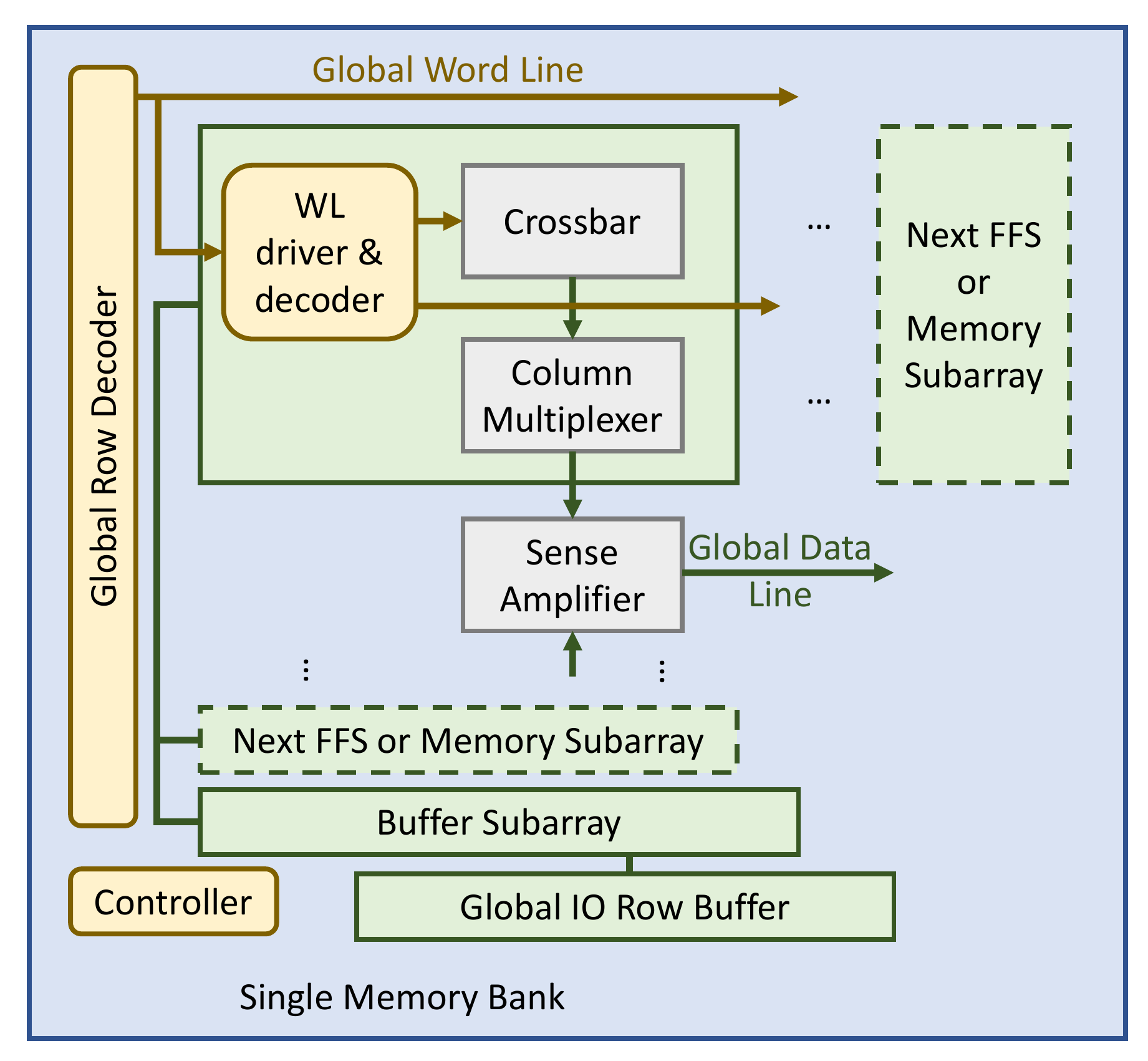}}
\caption{Illustration of the prime memory bank. Each FFS can operate in two modes, one is computation mode for MAC operation of DNNs, and the other is storage mode buffering and data preservation.}
\label{fig:PRIME}
\end{center}
\end{figure}

Different from ISAAC that never re-configures NVMs, the PRIME architecture~\cite{chi2016prime} uses a scheme where a portion of the NVM arrays can alternate between storage and compute units during runtime. As shown in Fig.~\ref{fig:PRIME}, the authors modify the standard wordline decoder and drivers (WDD), column multiplexers, and sense amplifiers so that they can better suit the RRAM-based crossbar arrays, and configure the storage banks into three different function units, memory subarrays (MS), full-function subarrays (FFS), and buffer subarrays (BS). The FFS is the key component that can alternate from memory to computational units. In the computation mode, FFS can perform VMM for DNNs, and in the storage mode, FFS buffers the intermediate data generated by VMMs. Similarly, the BS also acts as storage when FFS is not in computation mode. The sense amplifier is reconfigured to detect the higher precision analog value for computation compared to storage requirements so that that matrix multiplication can be performed. The modified column multiplexer executed analog substractions and nonlinear threshold functions. Although their implementation exerted a 60\% area overhead, the computation energy was saved by 94\% by reducing external memory accesses.



\section{Device and Circuit Non-idealities}\label{sect:devVar}

Although nvCiM can offer low latency and high energy efficiency, there are two major limitations of nvCiM, low data precision, and low device reliability. For the first issue, due to the limitation of the area and power budget, both the weight stored in the NVM devices and intermediate activation data can not be represented in a high precision manner. nvCiM DNN accelerators generally use data representations of four to eight bits~\cite{shafiee2016isaac, chi2016prime}. This problem is similar to the quantization problem of the traditional digital DNN accelerators and has been sufficiently discussed~\cite{han2015deep, wang2019haq}. However, the origin, simulation method, and mitigation approach of the reliability issue of nvCiM DNN accelerators are still open questions and are still receiving heated discussions. In this section, we introduce the origin of the reliability issue of nvCiM DNN accelerators with an example of RRAM devices. For STT and FeFET devices, the source of unreliability is similar but the significance and specific behavior of these noise sources are slightly different.

Various research about developing fault models for RRAM and other emerging NVM devices has been established. In this section, we focus on five noise sources that are directly related to the unreliability of nvCiM DNN accelerators: thermal noise, shot noise, random telegraph noise (RTN), programming errors, and endurance failures~\cite{feinberg2018making}.

\subsection{Thermal Noise}

Thermal noise is also known as Johnson-Nyquist noise. It is electronic noise caused by the thermal agitation of carriers and is a property of all passive devices. It happens regardless of whether a voltage is applied to the device. A well-established model for thermal noise is by placing a current source in parallel with the ideal target device. The current source is also known as the noise current and its magnitude is modeled by a Gaussian distribution with zero mean and a standard deviation of $\sqrt{\frac{4 K_B T \Delta f}{R}}$, where $K_B$ is the Boltzmann constant ($\approx 1.38\times 10^{-23} J/K$), $T$ is the temperature in Kelvins, $\delta f$ is the bandwidth of the signal measured, and $R$ is the resistance of the ideal target device. Thermal noise is a fundamental property of resistive circuit elements. From the model, we can observe that the only way to reduce thermal noise is to reduce the device temperature. To handle this source of noise, noise resilient architectures that can operate under thermal noise need to be devised.

\subsection{Shot Noise}
Shot noise is also a fundamental source of noise caused by the physical nature of electronic devices. This source of noise is called Poisson noise because it can be modeled by a Poisson process. The key cause of shot noise is the discrete nature of currents where electric currents actually consist of flows of discrete charges (\emph{e.g.}, electrons). When the number of electrons flowing through the device at a certain point of time fluctuates, a fluctuation of current through a device can be observed. This can affect the measurement accuracy when a detector is sensing the current flowing into it. Although shot noise is easy to be averaged out provided enough measurement time, devices working in high frequencies (\emph{e.g.}, nvCiM DNN accelerators) still suffer from such noise. As discussed above, a Poisson process is a more precise way of modeling shot noise, but this noise model is too complex when embedded in other models. A simpler model is a zero-mean Gaussian noise with a standard deviation of $\sqrt{2 q I \Delta f}$, where $q$ is the charge of an electron ($\approx 1.6 \times 10^{-19}$ coulombs), $I$ is the current flowing through the ideal target device, and $\Delta f$ is the bandwidth of the signal measured.

\subsection{Random Telegraph Noise}
\stolen{Random telegraph noise (RTN) exists in both CMOS and emerging NVM device circuits but is considered as a major cause of faults of emerging NVM devices~\cite{ielmini2010resistance}. RTN is also called burst noise and is caused by the charge carriers that are temporarily trapped inside the device, thus changing the effective resistance of the device. The result is a temporary and unexpected reduction in the resistance of a device at runtime. The trapping and untrapping of the charge carrier is modeled mathematically by means of the telegraph process, which is a Markovian continuous-time stochastic process that jumps discontinuously between two distinct values.}

\subsection{Programming Errors}
Programming errors refer to the difference between the actual device resistance and the target resistance due to the non-ideal configuration of the device. This is generally caused by both the process variations and temporal variations of each device. Affected by the former noise, when applied the programming voltage of the same magnitude and duration, the resistance of different instances of emerging NVM devices can be different. The latter leads to the fact even when applied to identical programming pulses, an NVM device can be programmed to different values in different trials of programming. A complex but effective way to mitigate this issue is to use a scheme called write-and-verify~\cite{alibart2012high, niu2012low, xu2013understanding}. The key operation is to iteratively apply a series of short pulses (write) and then check the difference between current and target resistance (verify), converging progressively on the target resistance. In deploying accelerators for Neural Network inference, this time-consuming progress is tolerable because once programmed, no more modifications to the resistance are needed during the entire life span of the accelerator. This scheme pulls down the programming error to less than 1\%. This 1\% of error can be modeled by a zero-mean Gaussian noise where the standard deviation is determined by the error upper bound of the write-and-verify process.

\subsection{Endurance and Retention}

\stolen{Endurance Failure is about the device being able to preserve their property after multiple times of write operations or and retention is about being able to read the desired data at a long period of time after programming. The endurance of emerging NVM devices varies widely based on the material properties and write mechanisms. The typical endurance for CMOS-based SRAM is $10^{16}$ which means typically, after this amount of write, the device would be stuck at a certain value, and writing it would be infeasible. The typical endurance for STT-MRAM is $10^{15}$, for FeFET is $10^5$ and for RRAM is $10^7$~\cite{feinberg2018making}. On the other hand, being able to read out the correct information when it is a long period of time after the device is programmed is also an important subject. This is called the retention issue. For simple CiM implementations like Memristive Boltzmann Machine, a typical worst-case lifetime is 1.5 years, but for nvCiM DNN accelerators, the system is more complex and the lifetime is shorter. To mitigate the effect of the endurance issue, researchers proposed a fault-tolerant online training method~\cite{xia2017fault} that maps the weight matrices stored in crossbars for computation around faults or endurance failures through a combination of neural network pruning and data remapping. This scheme increases the life of the neural network accelerator, allowing it to be used for training.}




\section{\todo{Impact of Device Variation on DNN Acceleration}}\label{Sect:Analysis}
\subsection{Model of Device Variation}

The source of device variations and their behaviors are introduced in Sect.~\ref{sect:devVar}, but modeling such device characteristics is not a simple task. A straightforward way is to abstract the behavior of different devices into circuit-level models~\cite{zhao2017investigation} and utilize circuit-level simulation tools (\emph{e.g.}, $SPICE$) to investigate the behavior of certain nvCiM DNN accelerators. However, because of the complexity of both neural network typologies and DNN accelerator architectures, building circuit-level models for nvCiM accelerators requires great human effort and needs to be modified each time a new type of accelerator architecture is proposed. Moreover, circuit-level models are computationally intensive. Using such models to simulate complex DNN accelerators requires considerable evaluation time and is not suitable during design phase explorations. Thus, a simple and effective model for the impact of device variations is needed.

One of the effective modeling methods is to model the device variation as a whole and use a Gaussian distribution to represent it~\cite{jiang2020device, feinberg2018making, gao2021bayesian, yan2020single}. Here we introduce one representative modeling method~\cite{gao2021bayesian} using Gaussian variables.

The NVM device electrical property, \emph{e.g.}, conductance, is subject to the combined effect of different variation sources as in Sect.~\ref{sect:devVar}. The actual conductance values g considering variations on n devices of a crossbar array can be written as:

\begin{equation}\label{eq:w1}
    \mathbf{g} = g_{0,n\times 1} + \Delta g_g + F(g_{0,n\times 1,\mathbf{r}})\approx \overline{g_{0,n\times 1}}+f(g_{0,n\times 1},\mathbf{r})
\end{equation}
\stolen{where $\overline{g_{0,n\times 1}} = g_{0,n\times 1} + \Delta g_g$ with $g_{0,n\times 1}$ denoting the expected conductance and $\Delta g_g$ denoting the global conductance variation as a constant for all the devices on the same die; $\mathbf{r}$ models the underlying spatially correlated and dynamic variations; $f(g_{0,n\times 1},\mathbf{r})$ is a function describing the dependence of variations on the expected conductance and can be approximated by $f(g_{0,n\times 1},\mathbf{r})$ due to the relatively small value of variations \emph{w.r.t.} the nominal values [11].}

\stolen{Since the mapped weights $\mathbf{w}$ are linearly related to conductance as $\mathbf{w} = c_1 \times \mathbf{g} + c_0$, where $c_1$ and $c_0$ are two constants, each weight $w_i$ represented by multiple devices can be modelled as a Gaussian variable:}

\begin{align}\label{eq:g_noise}
    w_i & = \mathcal{N}(u_{0,i},\Psi (u_{0,i})^2) \\
    & = \mathcal{N}(c_1 \overline{g_{0,i}} + c_0, c_1^2 f(\overline{g_{0,i}})^2(\Sigma^{m}_{k=1}\lambda_{i,k}^2+\lambda_{i,n}^2))
\end{align}

\subsection{Impact of Device Variation on DNN Outputs}\label{sec:changeMethod}

After finishing modeling the device variations, we can then investigate the impact of device variations on nvCiM DNN accelerators. A typical study is to evaluate such impact on an accelerator targeting image classification tasks~\cite{yan2021uncertainty}. In this section, we introduce the findings of the authors of~\cite{yan2021uncertainty}.

A starting point is understanding the effect of device variations on the output of a DNN model. The forward path of a DNN model can be viewed as a function of the input and the weight value of the model. Formally speaking, a DNN inference process can be defined as:

\begin{equation}\label{eq:NN}
    \mathbf{O} = F\ (W, \mathbf{I})
\end{equation}
where $F$ is the DNN architecture, $W$ is the DNN weights, $\mathbf{I}$ is the input vector, and $\mathbf{O}$ is the output vector. 

In classification tasks, the output vector $\mathbf{O}$ for each input (not batched) is a 1-D vector whose size is the number of possible classes. Each element of this vector represents the model's confidence that the input images should be classified into a certain class. Thus, the class with maximum value in $\mathbf{O}$ is what the model predicts to be the best choice for classification. During training, $\mathbf{O}$ is passed through a $Softmax$ function so that the confidence for each class is between 0 and 1 and the sum of confidences among different classes is 1. However, $Softmax$ is not necessary during DNN inference because it does not change the order of the values in $\mathbf{O}$. The final predicted class of $\mathbf{I}$ is calculated by $argmax(\mathbf{O})$, which is the index of the item in $\mathbf{O}$ that has the maximum value. As we focus on inference, the vanilla version of $\mathbf{O}$ before $Softmax$ is the key. 

Taking device variation into account, a model deployed on nvCiM DNN accelerators can be represented as:
\begin{equation}\label{eq:deploy}
    \mathbf{O}_{Dep} = F\ (W_{Dep},\ I) = F\ (\mathcal{N}(W_{Exp},\sigma),\ I) 
\end{equation}
where $W_{Dep}$ is the weight actually deployed on the accelerator and according to Eq.~\ref{eq:g_noise}, it can be modeled as a Gaussian variable whose mean is $W_{Exp}$, which is the trained value of the neural network to be deployed, and the standard deviation is $\sigma$, which can be calculated using Eq.~\ref{eq:g_noise}. $\mathbf{O}_{Dep}$ is the affected output.

One indicator of the effect of device variations on nvCiM accelerators is the difference in output. Formally speaking, we can define \emph{output change} as the difference between the output without device variation and the output value under the impact of device variation:

\begin{equation}\label{eq:change}
    \mathbf{O}_{Change} = F\ (W_{Exp},\ I)  - F\ (\mathcal{N}(W_{Exp},\sigma),\ I) 
\end{equation}
Note that $\mathbf{O}_{Change}$ is also a random variable.

In order to get a glance at the statistical behavior of $\mathbf{O}_{Change}$, according to the workflow introduced in Sect.~\ref{sec:changeMethod}, the authors train a LeNet model for the MNIST dataset~\cite{lecun1998gradient} to state-of-the-art accuracy. The authors then randomly choose one input image in the test dataset and sampled 10k different instances of noise. with this setup, the authors gathered 10k different $\mathbf{O}_{Change}$ vectors.

For MNIST, $\mathbf{O}_{Change}$ is a vector of 10, with each element representing the confidence of classifying the input image into one certain number digit. Because a high-dimensional vector is not a good choice for analytical study and visualization, each element of these vectors is visualized independently, so 10 instances of distribution data are collected.

\begin{figure}[h]
\begin{center}
\centerline{\includegraphics[trim=0 350 680 0, clip, width=0.6\linewidth] {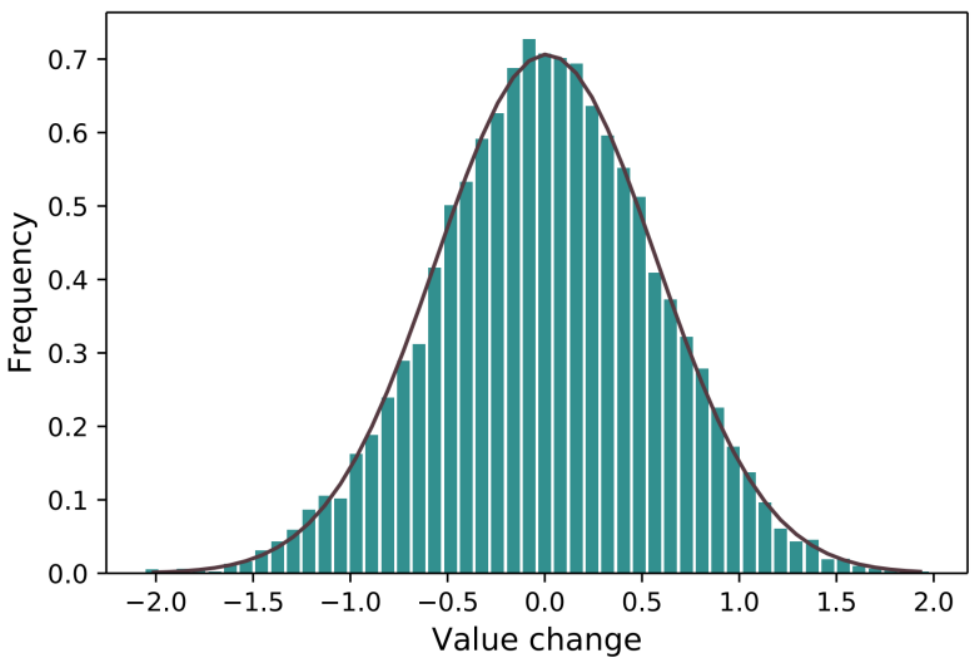}}
\caption{$\mathbf{O}_{Change}$ distribution of LeNet for MNIST. 10k $\mathbf{O}_{Change}$ vectors are gathered from one trained LeNet model affected by 10k different instances of weight values from $\sigma = 0.04$. This figure shows the distribution of the fist item of the gathered $\mathbf{O}_{Change}$ vectors. It is obvious that the visualized variable is Gaussian.}
\label{fig:MNIST_Output}
\vspace{-0.5cm}
\end{center}
\end{figure}

Each element of $\mathbf{O}_{Change}$ follows Gaussian distribution. To visualize this finding, the authors plot the histogram of the distribution of each element of $\mathbf{O}_{Change}$ vector and the corresponding Gaussian distribution that fits it. The visualization result for the first element of $\mathbf{O}_{Change}$ is shown in Fig.~\ref{fig:MNIST_Output}. It is obvious that the visualized variable is Gaussian.

This observation generalizes in various networks in various datasets. For the MNIST dataset, three models are analyzed: (1) LeNet and two-layer-multilayer perceptrons (2-layer-MLP) using (2) ReLU and (3) Sigmoid activation. For the CIFAR-10 dataset~\cite{krizhevsky2009learning}, the authors of~\cite{yan2021uncertainty} test four models: (1) a conventional floating-point CNN, (2) a quantized CNN, and two ResNets, (3) ResNet-56 and (4) ResNet-110. For each model, three different initializations are used to train three different sets of weights.

The authors of~\cite{yan2021uncertainty} collect all $\mathbf{O}_{Change}$ variables and find the closest Gaussian variable that fits each of them. To measure the similarity of $\mathbf{O}_{Change}$ and its Gaussian counterpart, two widely used standards: mean square error (MSE) and Chi-square ($\chi^2$) test are used. For variables with one element, MSE can be described as:

\begin{equation}\label{eq:MSE}
    MSE = \frac{1}{N} \sum^{N}_{i=1}(O_i - E_i)^2
\end{equation}
and $\chi^2$ test can be depicted as:
\begin{equation}\label{eq:chi2}
    \chi^2 = \sum^{N}_{i=1}\frac{(O_i - E_i)^2}{E_i}
\end{equation}
where $O_i$ and $E_i$ are the observed ($\mathbf{O}_{Change}$) and estimated (Gaussian) value of, normalized in the form of probability density, and $N$ is a user-defined granularity. Here $N = 100\ $ is used because it is precise enough when there is a total of 10k instances of $\mathbf{O}_{Change}$ data. The similarity of a vector is averaged out among all of its elements and the final similarity is also averaged out among all different initializations.

\begin{table}[h]
    \centering
    \begin{tabular}{lccc}
        \hline\noalign{\smallskip}
        Model & Dataset & $\chi^2$ ($10^{-2}$)& MSE ($10^{-4}$)\\
        \noalign{\smallskip}\svhline\noalign{\smallskip}
        MLP-ReLU    & MNIST     & 5.22 & 3.20\\
        MLP-Sigmoid & MNIST     & 5.81 & 2.20\\
        LeNet       & MNIST     & 4.59 & 2.67\\
        Float-Conv  & CIFAR-10  & 7.01 & 3.03\\
        Fixed-Conv  & CIFAR-10  & 6.79 & 2.74\\
        ResNet-56   & CIFAR-10  & 4.56 & 1.79\\
        ResNet-110  & CIFAR-10  & 4.81 & 2.01\\
        \hline\noalign{\smallskip}
    \end{tabular}
    \caption{The similarity of $\mathbf{O}_{Change}$ distribution and its Gaussian fit for different models. The $\chi^2$ test result and MSE between the $\mathbf{O}_{Change}$ and its Gaussian fit counterpart is presented. Both tests show that the $\mathbf{O}_{Change}$ is a multi-dimensional Gaussian variable \emph{w.r.t.} different instances of noise.}
    \label{tab:fit}
    \vspace{-0.5cm}
\end{table}

The similarity of $\mathbf{O}_{Change}$ distribution and its Gaussian fit for different models are shown in Table~\ref{tab:fit}. For each model tested, the average $\chi^2$ test results among different initializations are all below 0.1 and MSE are all below $10^{-3}$, which indicates we can have high confidence that $\mathbf{O}_{Change}$ distribution is Gaussian. Moreover, this observation is scalable because, for both extremely shallow (\emph{e.g.}, 2-layer MLP)  and very deep (RestNet-110) candidates, both errors do not increase. Thus this observation generalizes across different DNN models targeting classification tasks. With this conclusion, the authors of~\cite{yan2021uncertainty} claim that,

\noindent\textbf{with any independent and identically distributed Gaussian noise on weight, the output vector of the same input image follows a multi-dimensional Gaussian distribution\footnote{Note that each element of the output are deeply co-related, not independent} over different samples of noise.}

This claim is very strong and there is only empirical support for it. However, it is not counter-intuitive. The output of the first convolution layer is the summation of the multiplication result of deterministic inputs and Gaussianly distributed weights and is thus a summation of Gaussian distributions. The summation of Gaussian variables is also a Gaussian variable, so the output of the first layer is a Gaussian variable. After activation, the input of the second layer is a transformed Gaussian variable. After propagating through this layer, each output value is the sum of multiple multiplication results, and operands for each multiplication are both Gaussian variables. It is also worth noticing that, for the same layer, the standard deviation $\sigma$ for each noisy weight is the same. So the results of each multiplication are close to IID and with enough number of operands for this summation, the accumulated variable can be approximated by Gaussian variables. Thus, although the final output may not strictly be a Gaussian variable, a Gaussian approximation can be observed.




\section{Dealing with Device Non-idealities}

The majority of noise sources of nvCiM DNN accelerators are random noise that is difficult to eliminate during device production. Fortunately, there are opportunities from the accelerator architecture, DNN topology design, and DNN training aspects that can help to mitigate the effect of device variations. In this section, the authors introduce four different efforts from these three aspects

\subsection{Error Correction}

As discussed in Section.~\ref{sect:general_arch}, nvCiM accelerators process DNN models in a layer by layer manner and devise nvCiM processing units that consist of crossbar arrays and other peripheral digital blocks to perform matrix-vector multiplication and other key DNN operations including non-linear activation and pooling. From the accelerator architecture design aspect, it is a straightforward idea to equip nvCiM platforms with error correction abilities so that they can mitigate the effect of device variations.

In this section, we introduce one representative work~\cite{feinberg2018making} that uses error correction code to assist nvCiM computation. The authors use a group of arithmetic codes, named AN-codes~\cite{van2012introduction} for error correction. Arithmetic codes are a class of error correction codes (ECCs) that can preserve the result of arithmetic operations with noisy operands. AN-codes are a set of arithmetic codes that apply arithmetic weight to each operand so that it can maximize the arithmetic distance between codewords. An example of AN codes that utilizes residues is, for a given integer $K$ and operands $A$ and $B$, $K A + K B = K(A+B)$ and $(K A + K B) \% K \equiv 0$. The ECC units can detect and correct the error according to the residue.

\begin{figure}[h]
\begin{center}
\centerline{\includegraphics[trim=0 130 370 0, clip, width=0.6\linewidth] {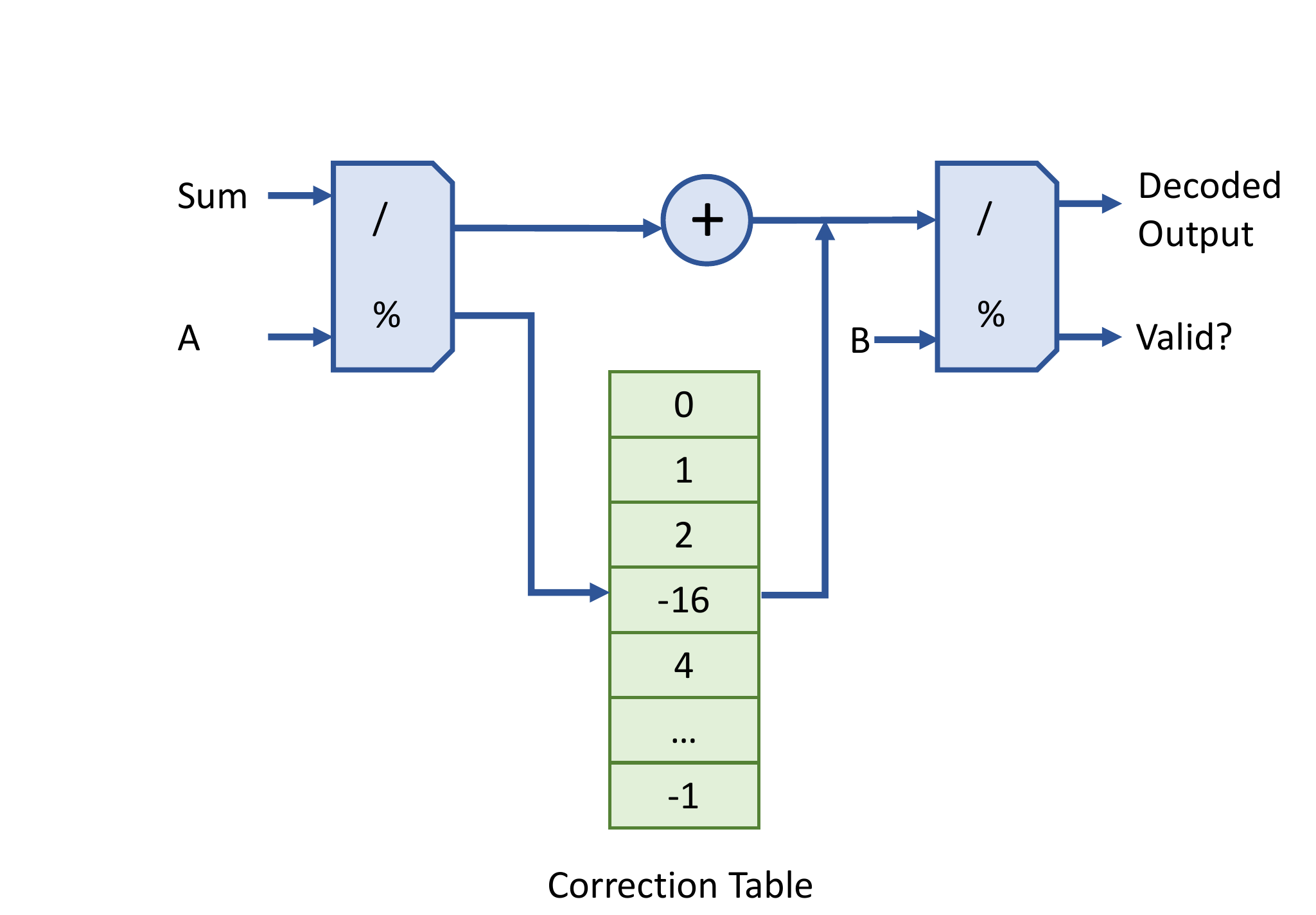}}
\caption{Illustration of error correction unit circuitry. This is a lookup table styled design.}
\vspace{-0.5cm}
\label{fig:ecc}
\end{center}
\end{figure}

\stolen{The error correction unit (ECU) in~\cite{van2012introduction} has three major components: two divide/residual units for the residual computation of A and B (one each), and a correction table that maps each residual to a syndrome. The output of the first divide/residual unit computes the integer division of the input by A and outputs the residual along with the quotient. The residual is used to index into the correction table, and the value read from the correction table is added to the result. This value is then fed into the second divide/residual unit where it is divided by B. The output of this unit is the final output of the error correction system and includes a flag indicating if the computation was in error. An illustration of ECU is shown in Fig.~\ref{fig:ecc}.}

\subsection{Identifying Robust Neural Architectures}


Some DNN topologies (neural architectures )are more robust than others against device variations. Finding these neural architectures is a viable way of mitigating the effect of device variations. Meanwhile, different neural architectures require different amounts of computation power and are thus with different inference latency and power consumption. Handcrafting a neural architecture that meets all design requirements is a challenging task. Fortunately, neural architecture search (NAS)~\cite{zoph2016neural, yan2021radars, yang2020co} is proposed to automatically find an optimal neural architecture in a designated design space using reinforcement learning-based algorithms. 

\stolen{In this section, we introduce NACIM~\cite{jiang2020device}, a device-circuit-architecture co-exploration framework that can automatically identify the best CiM neural accelerators from a design space including the device type, circuit topology, and neural architecture hyper-parameters. NACIM framework iteratively conducts explorations based on a reward function, which is suitable for reinforcement learning approaches or evolutionary algorithms. By configuring the parameters of the framework, designers can customize the optimization goals in terms of their demands. The authors model the effect of device variation by modeling the shot noise as a stuck-at-low or stuck-at-high fault, and the other noise sources as a whole to be a zero-mean additional Gaussian noise extracted from widely adopted models~\cite{zhao2017investigation} on the weight value. Experimental results show that the proposed NACIM framework can find the robust neural network with only 0.45 percent accuracy loss in the presence of device variation, compare with a 76.44 percent loss from the state-of-the-art NAS without considering device variation.}

\subsection{Training Robust DNNs}

DNN models with the same neural architecture but different weights can have very similar accuracy in ideal conditions but very different accuracy in the existence of device variations. Thus, finding proper weights that are robust against device variations in the training process is a desirable approach.


A straightforward way to find robust weights is to simulate the noisy forward path in the training process, \emph{i.e.}, in each iteration of training, the algorithm sample an instance of noise and add it to the weight in the forward and backpropagation path to calculate the gradient, then remove the noise when updating the weights.

This method is used in NACIM~\cite{jiang2020device} which is introduced before. For implementations in MNIST dataset~\cite{lecun1998gradient}, noise injection training can reduce the accuracy drop between the ideal model and model with device variations from 6\% to 0.5\%, and in CIFAR-10 dataset~\cite{krizhevsky2009learning}, noise injection training can reduce the accuracy drop from 76.44\% to 0.45\%.


A more advanced way to find robust weights is to seek help from Bayesian Neural Networks (BNN). \stolen{Bayesian neural network is known for a stochastic gradient variational Bayes framework applied to approximate posterior distributions over network parameters. By employing a prior distribution over the weight space, BNN allows us to introduce variation to the learning process to better fit the observations~\cite{gao2021bayesian}}

\stolen{A recent work~\cite{gao2021bayesian} uses BNN to improve the robustness of nvCiM accelerators. BNN requires a priori distribution and uses an estimated posterior to fit this distribution. The priori can be obtained from device variation models. These models are inferred from expert knowledge with the help of measurement, simulation, and historical data. The authors also use KL divergence as the regularization term to enforce the memristor variation structural characteristics.}

\stolen{Although the priori used in most recent works are carefully designed, they can still be imprecise or uncertain because of the measurement imperfectness and the ever-going evolution of emerging devices. To address this issue, the authors of~\cite{gao2021bayesian} propose a variance-adaptive priori to weigh the value of prior knowledge. The author modify the optimization objective of BNNs so that weights with larger values are more regularized by the priori, \emph{i.e.}, it allows placing heavier priorities on those critical weights (with higher magnitude) on crossbar arrays that are prone to receive more impact from device variations, thereby reducing oscillations in convergence for more efficient training. Finally, to prevent the over-amplification of variation during training, the authors add an additional regularization term using L2 norm loss. In CIFAR-10 dataset, this proposed method is able to reduce the accuracy drop from 45.7\% to 0.3\%.}

Although these two methods are effective in terms of mitigating the effect of device variations on nvCiM accelerators, they require much more training iterations to converge compared with traditional training methods. In the MNIST dataset, both methods require at least 10X more iterations of training to reach a similar accuracy as the traditional training method~\cite{jiang2020device}.

\section{Conclusions}

Computing-in-memory with emerging non-volatile devices (nvCiM) is a great candidate for efficient DNN acceleration because of its unique architecture that breaks the memory wall. However, it suffers from unreliability issues, especially the device variation issues of emerging NV devices.  Understanding the property of emerging NV devices and the general architecture of nvCiM DNN accelerators helps to better model the effect of device unreliability circuit and application level. The modeling of unreliability also helps in mitigating the impact of device variations. The representative ways of mitigation include the adoption of ECC in the architecture and finding neural network topologies and training DNN weights that are more robust against device variations.

\bibliographystyle{spmpsci}
\bibliography{M5_References}
\end{document}